\documentclass[%
 reprint,
 amsmath,amssymb,
 aps,
]{revtex4-1}
\usepackage{graphics}
\usepackage{graphicx}
\usepackage{amssymb}
\usepackage{amsmath}
\usepackage{color}
\usepackage{bm}

\usepackage{amsmath}

\begin{document}


\title{Comment on: Multipartite entanglement in four-qubit graph state
}

\author{Saeed Haddadi}
\email{haddadi@physicist.net}

\affiliation{%
Department of Physics, Payame Noor University, P.O. Box 19395-3697, Tehran, Iran.
}

\begin{abstract}
\textsf{\textbf{Abstract.}} The following comment is based on an article by M. Jafarpour and L. Assadi [Eur. Phys. J. D {\bf 70}, 62 (2016), doi:\href{http://link.springer.com/article/10.1140/epjd/e2016-60555-5}{10.1140/epjd/e2016-60555-5}] which with an exploitation of Scott measure (or generalized Meyer-Wallach measure) the entanglement quantity of four-qubit graph states has been calculated. We are to reveal that the Scott measure ($Q_{m}$)  nominates limits for $m$  which would prevent us from calculating $Q_{3}$ in four-qubit system. Incidentally in a counterexample we will confirm as it was recently concluded in the mentioned article, the $Q_{2}$ quantity is not necessarily always greater than $Q_{3}$.
\begin{description}

\item[PACS numbers]
03.65. Ud, 03.65. Mn, 03.67.-a.

\end{description}
\end{abstract}

\pacs{Valid PACS appear here}
\maketitle

\begin{figure}
  \centering
  \includegraphics[width=0.90in]{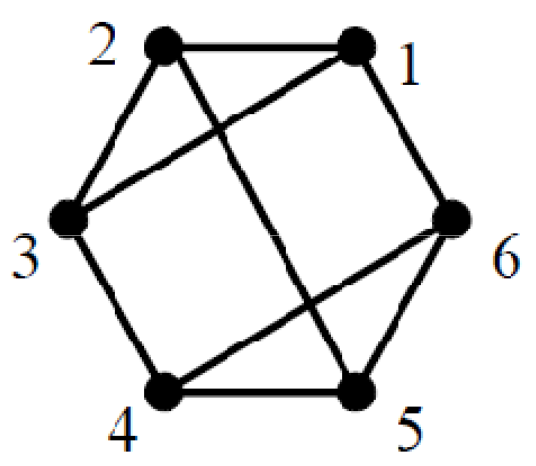}
  \caption{Six-qubit graph (as an example).}\label{Figure. 1}
\end{figure}

Recently, M. Jafarpour and L. Assadi \cite{Nielsen01} based on Scott measure have calculated the entanglement quantity in non-trivial four-qubit graphs. Scott studied various interesting aspects of $N$-qubit entanglement measures given by \cite{Love02,Scott03}
\begin{equation}
\label{E1}
Q_{m}(|\psi\rangle)=\binom{N}{m}^{-1}\sum_{|S|=m}\frac{2^{m}}{2^{m}-1}(1-\mathrm{Tr}[\rho_{S}^{2}]),
\end{equation}
where $S\subset \{1,\cdots, N\} $ and $\rho_{S}$ = $\mathrm{Tr}_{\acute{S}}(|\psi\rangle\langle\psi|)$ is the reduced density matrix for $S$ qubits after tracing out the rest. Also $m=1,\cdots,\lfloor\frac{N}{2}\rfloor$ and $\lfloor\frac{N}{2}\rfloor$ is the integer part of $\frac{N}{2}$. The $Q_{m}$ quantities $(0\leq Q_{m} \leq 1)$ correspond to the average entanglement between subsystems that consists $m$ qubits and the remaining $N-m$ qubits \cite{Borras04}. Meanwhile, $Q_{m}$ is invariant under local unitary transformations (LU), non-incremental on average under local operations and classical communication (LOCC). Hence on account of four-qubit system, we are only authorized to merely calculate $Q_{1}$ and $Q_{2}$. We have obtained $Q_{1}=1$ for all non-trivial four-qubit graphs (No. 1-41). Whereas the authors have calculated $Q_{3}$ in Table 1, leading to an incorrect result.
Thus Section $6-d$ (Conclusions and discussion) leads to $Q_{2}$ being always greater than $Q_{3}$ in all the graph states. We will rectify in a counterexample their achieved result is incorrect in general. To clarify, take graph $G_{\star}$ for example, which is plotted in Fig. \ref{Figure. 1}. The graph state corresponding to graph $G_{\star}$ is as followed
\begin{align}
|G_{\star}\rangle&=\frac{1}{8}(|000\rangle|\phi_{1}\rangle+|001\rangle|\phi_{2}\rangle+|010\rangle|\phi_{3}\rangle+|011\rangle|\phi_{4}\rangle\nonumber\\
&\quad\ +|100\rangle|\phi_{5}\rangle+|101\rangle|\phi_{6}\rangle+|110\rangle|\phi_{7}\rangle+|111\rangle|\phi_{8}\rangle).\nonumber\\
\end{align}

Where

\begin{align}
|\phi_{1}\rangle&=\{|000\rangle+|001\rangle+|010\rangle-|011\rangle\nonumber\\
&\,\,\:+|100\rangle-|101\rangle-|110\rangle-|111\rangle\},\nonumber\\
|\phi_{2}\rangle&=\{|000\rangle+|001\rangle+|010\rangle-|011\rangle\nonumber\\
&\,\,\:-|100\rangle+|101\rangle+|110\rangle+|111\rangle\},\nonumber\\
|\phi_{3}\rangle&=\{|000\rangle+|001\rangle-|010\rangle+|011\rangle\nonumber\\
&\,\,\:+|100\rangle-|101\rangle+|110\rangle+|111\rangle\},\nonumber\\
|\phi_{4}\rangle&=\{|000\rangle-|001\rangle+|010\rangle-|011\rangle\nonumber\\
&\,\,\:+|100\rangle-|101\rangle+|110\rangle+|111\rangle\},\nonumber\\
|\phi_{5}\rangle&=\{|000\rangle-|001\rangle+|010\rangle+|011\rangle\nonumber\\
&\,\,\:+|100\rangle+|101\rangle-|110\rangle+|111\rangle\},\nonumber\\
|\phi_{6}\rangle&=\{|001\rangle-|000\rangle-|010\rangle-|011\rangle\nonumber\\
&\,\,\:+|100\rangle+|101\rangle-|110\rangle+|111\rangle\},\nonumber\\
|\phi_{7}\rangle&=\{|001\rangle-|000\rangle+|010\rangle+|011\rangle\nonumber\\
&\,\,\:-|100\rangle-|101\rangle-|110\rangle+|111\rangle\},\nonumber\\
|\phi_{8}\rangle&=\{|001\rangle-|000\rangle+|010\rangle+|011\rangle\nonumber\\
&\,\,\:+|100\rangle+|101\rangle+|110\rangle-|111\rangle\}.
\end{align}

For six-qubit graphs the authorized $m$ is equivalent to 1, 2 and 3 $(m=1,2,3)$. Therefore $\mathrm{Tr}\rho_{i}^{2}$, $\mathrm{Tr}\rho_{ij}^{2}$ and $\mathrm{Tr}\rho_{ijk}^{2}$ are calculated as following

\begin{align}
& \mathrm{Tr}[\rho_{i}^{2}]=\frac{1}{2},\quad\ i\in\ S,
\end{align}

\begin{align}
& \mathrm{Tr}[\rho_{ij}^{2}]=\frac{1}{4},\quad\ i < j\in\ S,
\end{align}

\begin{align}
& \mathrm{Tr}[\rho_{ijk}^{2}]=\frac{1}{8},\quad\ i < j < k\in\ S.
\end{align}
In this calculation the final result will be
\begin{align}
Q_{1}(|G_{\star}\rangle)=Q_{2}(|G_{\star}\rangle)=Q_{3}(|G_{\star}\rangle)=1.
\end{align}

In conclusion, the analysis above shows that we are only authorized to merely calculate $Q_{1}$ and $Q_{2}$ for four-qubit system. Accordingly,  the calculation of $Q_{3}$ given by M. Jafarpour and L. Assadi \cite{Nielsen01} is unauthorized and ineffective.  Moreover, we note that $Q_{2}$ is not necessarily greater than $Q_{3}$  in all the graph states (or in general) but in some cases $Q_{2}$ is equal to $Q_{3}$.  M. Jafarpour and L. Assadi study four-qubit graph states for which they can choose to study ($Q_{1}$ and $Q_{2}$) or ($Q_{2}$ and $Q_{3}$), since in fact $Q_{1}$ and $Q_{3}$ are proportional to each other by a numerical factor (as seen from Eq. (\ref{E1})). In fact for four-qubit system $Q_{2}$ refers to 2-2 partitions in the graph state and $Q_{1}$ or $Q_{3}$ both refer to 3-1 partitions of the same graph. If we consider the case $Q_{1}$, 3-1 partition presents a stronger entanglement than a 2-2 partition in non-trivial four-qubit graphs (Unlike a result of the aforementioned article).

%
%

\end{document}